\documentclass[oneside]{article}
\usepackage[T1]{fontenc}
\usepackage{authblk}
\usepackage{float}
\usepackage{amsmath}
\usepackage{graphicx}
\usepackage{xfrac}
\usepackage[english]{babel}
\usepackage{lmodern}
\usepackage[T1]{fontenc}
\usepackage[latin9]{inputenc}

\usepackage{csquotes}

\usepackage{longtable}

\usepackage{mathtools}
\usepackage{graphicx}
\usepackage{esint}
\usepackage[unicode=true,
 bookmarks=false,
breaklinks=false,pdfborder={0 0 1},backref=false,colorlinks=false]
 {hyperref}
\usepackage[a4paper, total={210mm,297mm},margin=2.0cm]{geometry}

\title{Towards a self-consistent Boltzmann's kinetic model of fluid turbulence} 
\author[1,2,3]{Sauro Succi}

\affil[1]{Center for Life Nano Science@La Sapienza, Istituto Italiano di Tecnologia, 00161 Roma, Italy}
\affil[2]{Istituto per le Applicazioni del Calcolo CNR, via dei Taurini 19, Rome, Italy}
\affil[3]{Institute for Applied Computational Science, John A. Paulson School of Engineering and Applied Sciences, Harvard University, Cambridge, USA}

\begin{document}

\maketitle

\begin{abstract}
A closure for the effective relaxation time of the Boltzmann-BGK kinetic 
equation for fluid turbulence is presented, based on a double-averaging 
procedure over both kinetic and turbulent fluctuations.  
The resulting effective relaxation time appears to agree with 
values obtained via a renormalization group treatment of 
the Navier-Stokes equation only at low values of $k/T$, the ratio of 
turbulent kinetic energy to fluid temperature.
For $k/T>0.1$ the kinetic treatment delivers a significantly 
longer effective relaxation time. 
\end{abstract}

\section{Introduction}

The basic equations of fluid mechanics are known for two 
centuries, and yet, fluid turbulence keeps standing as one of the most 
challenging and compelling problems in modern science, holding back 
progress across many fluid-related disciplines and application 
in science and engineering \cite{FRISCH}.

Computer simulations have moved great lengths in the direction of 
unraveling the complexity of fluid flows, and yet, even most powerful 
foreseeable (non-quantum) computers fall short of providing access to 
the direct simulation of most flows of practical
interest, such as a full car or airplane, not to mention geophysics, 
astrophysics and cosmology.

Hence, major efforts are devoted to the task of devising the effects of
the small unresolved scales on the large and resolved ones, an art 
known as turbulence modeling (TM) \cite{TM}.

A central idea of TM is the notion of eddy viscosity, whereby the collective
degrees of freedom of turbulence ("eddies") are treated in full analogy 
with molecules in kinetic theory \cite{BOUSSI}.
This means that the large eddies experience a sort of Brownian motion due
to the erratic collisions with small eddies, leading to 
the notion of "eddy viscosity".
This concept has proved extremely valuable but suffers 
of a basic flaw, namely the assumption of scale separation between short and large eddies.
While suitable for molecules, such scale-separation fails 
for fluid turbulence, owing to the continuum spectrum of turbulent eddies.

Thus, what one needs is a theoretical and computational framework capable
of dealing with the non-perturbative aspects of eddy interactions across
scales of motion, and most notably with the interactions 
between eddies of nearby size.

Kinetic theory is ideally positioned to offer such a framework, since 
the Boltzmann equation requires no separation between micro and 
macroscopic scales.  
In more technical terms, it applies at any value of the 
Knudsen number, leading 
to hydrodynamics in the limit of zero Knudsen numbers.

However, the Boltzmann equation has been traditionally 
disregarded by the turbulence
community mainly account of its computational 
complexity: why solving a $(6+1)$-dimensional
equation to attack a problem which lives in $(3+1)-d$ only?

Over the last decade, this position has been revisited thanks to the vigorous
development of the lattice Boltzmann (LB) method, which is based on a 
minimal Boltzmann equation, living in a discrete and uniform lattice \cite{LB}.

More precisely, turbulence models based on suitable 
extensions of the LB have been
developed and applied to a variety of ideal and real-life 
turbulent flows \cite{JFM1,JFM2,SCI}.

Notwithstanding their practical success, such an approach has been criticised
on account of lack of self-consistency, namely the fact of coupling the
LB to a macroscopic equations for the fluctuating kinetic energy, which is derived
from coarse-graining of the macroscopic fluid equations.

In this paper we (partially) mend this weakness by showing that 
a very simple kinetic closure leads to a formulation very similar to
the one obtained in the macroscopic  approach.  

\section{Kinetic equation for turbulent eddies}

The main idea of the kinetic approach to fluid turbulence is 
to coarse-grain the Boltzmann
kinetic equation {\it before} taking the Chapman-Enskog limit 
from the kinetic to the hydrodynamic level \cite{PRE99,JFM1,JFM2}. 
This contrasts with the standard hydrodynamic approach , which 
consists in coarse-graining the Navier-Stokes equations. 

Symbolically:
\begin{equation}
U_h (x;t) =  {\mathcal P}_x \cdot {\mathcal P}_v \cdot f(x,v;t)
\end{equation}
for the hydrodynamic approach, versus the kinetic one: 
\begin{equation}
U_k (x;t)= {\mathcal P}_v \cdot {\mathcal P}_x \cdot f(x,v;t).
\end{equation}

In the above, ${\mathcal P}_x$ denotes a space-filtering projection 
operator, while ${\mathcal P}_v$ denotes the
velocity projector associated with Chapman-Enskog asymptotics.
In full generality, we expect the two projectors to commute 
only whenever the coarse-grained 
mean-free path $\Lambda_m$ remains sufficiently smaller than the 
lattice spacing $\Delta$, i.e.
$$
Kn_{\Delta} = \Delta/\Lambda_m\ll 1
$$
This condition is tantamount to assuming a scale-separation between the resolved and 
unresolved eddies, an assumption that, while inevitable in the hydrodynamic picture, is
guaranteed to fail for eddies in the vicinity of the lattice cutoff $\Delta$.
This very plain observation highlights the potential of the kinetic approach 
to deliver a genuine new class of turbulence models, free of scale-separation 
assumptions, hence more suitable to handle strong non-equilibrium 
turbulence, as it typically occurs in the vicinity of solid boundaries. 

To implement the above program, we start from the Boltzmann equation 
in single-time relaxation form (BGK) \cite{BGK}:

$$
\partial f = \frac{1}{\tau_0} (f^{eq}-f)
$$

where $\partial \equiv \partial_t + v \partial_x$ is the Lagrangian derivative
along the molecular velocity (streaming operator) 
and $\tau_0$ is the molecular relaxation time .

In the above $f(x,v;t)$ is the probability of finding a molecule 
at position $x$ with velocity $v$ at time $t$ and
$f^{eq}$ is a local Maxwellian at temperature $T$ and average
fluid velocity $u(x,t)$. Vector indices are relaxed for simplicity.

The basic coarse-grained kinetic quantities are defined as follows:
\begin{eqnarray}
F \equiv <f>,\\ 
F^{eq} \equiv  f^{eq}(<f>),\\ 
\delta F^{eq} \equiv <f^{eq}(f)>-f^{eq}(<f>)
\end{eqnarray}
the latter being the contribution of the nonlinear turbulent fluctuations
$u'=u-U$, $U=<u>$, $<u'>=0$, to the coarse-grained equilibrium, brackets denoting
ensemble averaging over turbulent fluctuations.

In the renormalization group language, one would like to understand 
how the Boltzmann-BGK equation transforms under the following rescaling:
\begin{eqnarray}
x \to bx,\; t \to bt,\; u \to u,\; v \to v,\;
\tau \to b^{\alpha} \tau
\end{eqnarray}
where $\tau_R \equiv b^{\alpha} \tau$ is the renormalised relaxation time, 
accounting for coarse-grained nonlinear contributions.
In particular, if $\alpha>1$, the lattice Knudsen number 
increases in the large-scale limit, so that 
space filtering and the Chapman-Enskog expansion do not necessarily 
commute, thus leading to a potentially new class of kinetic turbomodels 
with no hydrodynamic counterpart \cite{PRE99,ANSU}.

Formal coarse graining (filtering) of the kinetic equation delivers:
\begin{equation}
\partial F =-\frac{1}{\tau_0} (\delta F^{neq}+\delta F^{eq})
\end{equation}
where we $\delta F^{neq} \equiv F-F^{eq}$ is the coarse-grained
non-equilibrium and $\delta F^{eq}$ is the contribution from the
nonlinear fluctuations of the fine-grained local equilibrium $f^{eq}$.

From the above, we {\it formally} derive the following 
"renormalised" relaxation time (RRT):
\begin{equation}
\label{TAUR}
\frac{\tau_R}{\tau_0} = \frac{1}{1-R}
\end{equation}
where we have set
\begin{equation}
R = \delta F^{eq}/\delta F^{neq}
\end{equation}
In other words, the RRT depends only on the ratio between the 
turbulent and kinetic fluctuations.
Note that in the absence of coarse-graining, $\delta F^{eq}=0$, 
and $\tau_R \to \tau_0$, as it should be by mere consistency.

Three distinguished regimes are apparent.

{\it 1) Contraction regime ($R<0$)}: 
the turbulent fluctuations carry an opposite sign as compared to 
the kinetic ones, so that the renormalized relaxation time is shorter 
than the bare one (contraction).
This is an unlikely situation, which may eventually occur for supersmooth 
regimes, in which the velocity fluctuations scale superlinearly with 
the size of the eddies, $\delta u(l) \propto l^{1+\alpha}$, $\alpha>0$,
so that $\tau(l) = l/\delta u(l) \sim l^{-\alpha}$.

{\it 2) Dilatation regime ($0<R<1$)}: 
the turbulent fluctuations carry the same sign as the non-equilibrium
ones, but they are smaller in amplitude.
Consequently, the RRT exceeds the bare relaxation time and 
diverges in the limit $R \to 1$. 

The relation (\ref{TAUR}) shows that largest RRT's arise in connection
to turbulent fluctuations getting close to the kinetic ones, yet smaller. 
The physical interpretation is that in the range $0<R<1$, the 
renormalised equilibrium $<f^{eq}>$ gets closer
to the actual coarse-grained distribution $<f>$ than 
the bare coarse-grained equilibrium $f^{eq}(<f>)$, which is tantamount 
to a dilatation of the renormalised relaxation time (RRT). 
In this regime scale separation breaks-down and the kinetic approach 
is expected to deliver genuinely new results.  

{\it Unstable regime ($R>1$)}: the turbulent fluctuations still
carry the same sign as kinetic ones, but now they are larger in amplitude. 
As a result, $\tau_R$ becomes formally negative, which hints at an 
instability, since it is as if in order to attain the equilibrium, the system 
should go back in time, which manifestly it cannot do.
While we are in no position to assess the realizability of such 
regime, we simply observe that occasional instabilities are 
definitely part of the picture in the case of 
non-equilibrium turbulence (gusts of intermittency).

Leaving a more detailed inspection of these three regimes 
to a future publication, we next proceed to a quantitative assessment
of both turbulent and kinetic fluctuations.

\section{Turbulent fluctuations: coarse-grained equilibria}

Under the assumption of ergodicity, coarse-graining can be formulated 
as a space-time filter of the form:
$$
F(x,v,t) \equiv <f> = \int K(x-x',t-t') f^{eq}(x',v,t') dx' dt'
$$
In practice, this all but a convenient procedure, for it requires 
homogeneous directions to average upon, which
are hardly available in real-life geometries \cite{LES}.

Hence, we take a different route, first developed by Yakhot \cite{YAK,IOP}, which 
replaces spacetime averaging with ensemble averaging in kinetic space.
More precisely, one decomposes the molecular velocity as follows:
$$
v=U+u'+v' \equiv V+u'
$$
where $u'$ are the turbulent fluctuations, $v'$ are 
the kinetic ones and we have set $V=U+v'$.

By ergodicity, we assume that the filtering in space can be replaced by 
a (functional) average over the turbulent fluctuations, namely:
$$
F(U+v') = \int f(U+v'+u') P(u'|U) du'
$$
Next we make the plausible assumption that the one-point velocity 
fluctuations are gaussian distributed with variance $k=<u'^2>/2$, 
the "turbulent temperature", i.e., in $d$ spatial dimensions:
\begin{equation}
P(u'|U) = (2 \pi k)^{-d/2} e^{-u'^2/2k}
\end{equation}

Note that for one-point fluctuations in homogenous turbulence, this assumption 
is a pretty safe one.

Since the local equilibrium is gaussian, and so is the one-point distribution, the
above integral can be performed  analytically, to deliver a 
"Doppler" shifted gaussian with temperature
$T \to T+k$, where $k=<u'^2>/2$ is the turbulent kinetic energy 
(we have set $k_B=1$ and unit density since 
we deal with incompressible flows).

Thus, the coarse-grained BGK equation reads as follows:
\begin{equation}
\label{EQF}
\partial F = \frac{1}{\tau_0} (F^{eq}_{k+T}-F)
\end{equation}
which looks exactly the same as the original one, only 
with a Doppler shifted equilibrium.

The equation is not closed, though, as it requires the dynamics of 
the turbulent kinetic energy $k$.

This can be derived by multiplying the BGK equation by 
$u'^2/2$ and performing the double integration upon $v'$ and $u'$, namely
\begin{equation}
k(x,t)= <<u'^2>> = \int F(U+u'+v';x,t) P(u'|U)  u'^2 du' dv'.
\end{equation}

The resulting equation is \cite{IOP}:
\begin{equation}
\label{EQK}
\partial_t k + U \partial_x k + \partial_x <u'^3> = -\frac{1}{\tau_0} (k-k_{eq}) 
\end{equation}
Performing the algebra and setting cross-correlation terms $<u'u'v'>$ to zero (true only at equilibrium)
we obtain $k_{eq} \sim kT/(k+T)=k/(1+k/T)$. This interpolates between $k$ 
in the limit $k/T<<1$ and $T$ in the opposite limit $k/T>>1$, the former being
usually the relevant case for fluid turbulence.

The skewness term $<u'^3>$, requires a non-equilibrium closure, examples
of which can be found in \cite{YAK}. 

Here, however, we wish to pursue a different goal, namely, in line 
with RG ideas, leave the coarse-grained equilibria 
invariant and formulate a kinetic closure for the
the renormalized relation time $\tau_R$.
 
Based on the above, by definition:
\begin{equation}
\label{DEQ}
\phi(c,\kappa) \equiv \frac{\delta F^{eq}}{F^{eq}}=
\frac{(T+k)^{-3/2} e^{-w^2/2(T+k)}}{T^{-3/2} e^{-w^2/2T}}-1
\end{equation}
where $w=v-U$ is the coarse-grained peculiar speed.

A simple rearrangement yields:
\begin{equation}
\label{DEQ2}
\phi(c,\kappa)= \frac{e^{\frac{c^2}{2} 
\frac{\kappa}{1+\kappa}}}{(1+\kappa)^{3/2}}-1
\end{equation}
where we have defined $\kappa \equiv k/T$ and $c \equiv w/T^{1/2}$.

It can be readily checked that such quantity hardly exceeds 
$1$, other than for superthermal excitations with $c \gg 1$.
Since such superthermal excitations are exponentially 
suppressed in the molecular fluid, we conclude
that $\delta F^{eq}/F^{eq}$ is generally well below $1$ (see Figure 1).
\begin{figure}
\centering
\label{FIG1}
\includegraphics[scale=0.3]{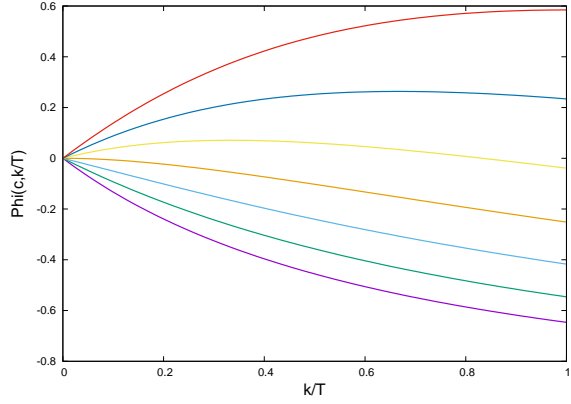}
\caption{The ratio of turbulent fluctuations to the bare coarse-grained
equilibrium as a function of the parameter $k/T$ for different values of
the peculiar speed $c^2=0,1,2,3,4,5,6$, top to bottom.} 
\end{figure}

For instance, to order $\kappa^2$, we obtain:
$$
<\delta F^{eq}/F^{eq}> = (1-(3/2) \kappa + (9/4) \kappa^2) 
(1+\frac{<c^2>}{2} \frac{\kappa}{1+\kappa}  + 
\frac{<c^4>}{8} \frac{\kappa^2}{(1+\kappa)^2}) + O(\kappa^4) 
$$
where brackets denote integration upon the peculiar velocity.
By recalling that in $d=3$, $<c^2>=3$, and $<c^4>=9$, we obtain:

\begin{equation}
\label{PHIEQ0}
<\delta F^{eq}/F^{eq}> = -\frac{3}{4} \kappa^2 +O(\kappa^3)
\end{equation}
Since $\kappa \sim 0.1$, this is of the order of $0.01$, thus 
showing that the scale-separation breaking
regime $R \to 1$ is attained through 
nonequilibrium heterogeneity effects sligthly below such value.

A crucial caveat must be pointed out: the relation $<c^2>=3$ and 
$<c^4>=9$ only hold at equilibrium, which means that performing the average with
the actual distribution delivers a linear contribution
$(<c^2>-3)\kappa/2$.
The ratio of non-equilibrium to equilibrium distribution scales
like the Knudsen number, $Kn \equiv \tau_R \partial \sim \tau_R/\tau_{tur}$,
where $\tau_{tur}$ is a typical turbulent time scale to be discussed 
in the next section.

In view of such observation, we finally write
\begin{equation}
\label{PHIEQ}
\Phi \equiv <\delta F^{eq}/F^{eq}> = C_1 \kappa -\frac{3}{4} \kappa^2 +O(\kappa^3)
\end{equation}
where $C_1 \sim Kn$.
Hence, the linear contribution in $k/T$ is a genuine non-equilibrium effect.

\section{Coarse-grained non-equilibrium}

The denominator of eq (\ref{TAUR}), can be computed by solving 
the coarse-grained BGK equation in the form $(1+\tau_R \partial) F = F^{eq}$. 
This delivers:
\begin{equation}
\label{FNEQ}
\delta F^{neq} = -\lbrace \frac{r z}{1+r z} \rbrace F^{eq}
\end{equation}
where we have defined 
$r = \tau_R/\tau_0$ and $z \equiv \tau_0 \partial$ (the Knudsen operator).

Inserting (\ref{FNEQ}) in (\ref{TAUR}), we obtain a self-consistent 
operator equation for the renormalized relaxation time $\tau_R$.

This is a fully non-local operator equation, since $z$ involves the
streaming operator $\partial$, but we shall treat it as an ordinary number
by invoking the {\it correspondence rule} $\partial = 1/\tau_{tur}$, subscript 
"tur" standing for "turbulent".

By letting $\theta \equiv \tau_{tur}/\tau_0$, the relation (\ref{FNEQ}) simplifies to

\begin{equation}
\label{KIN0}
r-1=-(1+\theta)\frac{\phi}{1+\phi} 
\end{equation}
where we remind that $\phi \equiv \delta F^{eq}/F^{eq}$.
Integration upon the peculiar velocity provides
\begin{equation}
\label{KIN}
r-1=-(1+\theta)\frac{\Phi}{1+\Phi} 
\end{equation}
where $\Phi = <\phi>$ and we have made the assumption $\phi \ll 1$.

The next task then is to pin down a concrete expression for 
the unknown timescale $\tau_{tur}$.

A natural correspondence rule is as follows:

\begin{equation}
\label{TAUTUR}
\frac{1}{\tau_{tur}}= {(\frac{1}{\tau_l^2}+ \frac{1}{\tau_s^2})}^{1/2}
\end{equation}
where 
$\tau_l = k/\epsilon$ is the local timescale of homogeneous
turbulence and $\tau_s = 1/S$ is the inhomogeneity scale, $S=\partial_r u$ 
being the shear rate. The ratio of the two, often denoted as strain parameter,
$\eta = kS/\epsilon$ is a measure of non-equilibrium between eddies of
different size, $\eta=0$ denoting the equilibrium case (no strain). 

The above correspondence rule is tantamount to postulating that the 
time derivative in the streaming operator 
contributes a term $1/\tau_l$, where $\tau_l$ is a
typical time scale of homogeneous turbulence, namely $\tau_l \sim k/\dot k$.

Likewise, it appears plausible to assume that the spatial derivative 
$v \partial_x$ contributes a term of order $1/\tau_s = S$ where 
$S \sim \partial_x U$ is the large-scale shear.
The square is for the sake of positivity, but any higher even power would do.

Putting together the expressions (\ref{PHIEQ}) and (\ref{KIN}), we
we arrive at the following expression (in the limit $k/T \ll 1$):
\begin{equation}
\label{RRT0}
\frac{\tau_R}{\tau_0} - 1 = 
\frac{C_1 \kappa + C_{2} \kappa^2}{1+C_1 \kappa +C_{2} \kappa^2} \; 
[1+\frac{\tau_l/\tau_0}{(1+\frac{\tau_l^2}{\tau_s^2})^{1/2}}]
\end{equation}
where $C_1 \sim Kn$ and $C_{2} \sim 3/4$.

Noting that the coefficient $C_1$ is proportional to the
Knudsen number, hence depends on $\tau_R$ itself, we rearrange the above
expression in the following form:
\begin{equation}
\label{RRT}
\frac{\tau_R}{\tau_0}-1 = 
\frac{K_1 \kappa + K_{2} \kappa^2}{1+K_3 \kappa + O(\kappa^2)}\; 
\frac{\tau_l/\tau_0}{(1+\frac{\tau_l^2}{\tau_s^2})^{1/2}}
\end{equation}
where $K_1 = \tau_0/\tau_{tur}$, $K_2=3/4$, $K_3=1+\tau_0/\tau_{tur}$ and
we have made the the assumption $\tau_0/\tau_{tur} \ll 1$.

This is the main result of this paper, in that it provides
a kinetic closure for the RRT`in terms of the ratio $k/T$ and
the turbulent timescale $\tau_{tur}$.

\section{Comparison with Yakhot-Orszag renormalization group treatment}

The above treatment suggest a general expression for the RRT, namely
\begin{equation}
r-1= \Psi^{eq}(k/T) \; \Psi^{neq} (\frac{\tau_l}{\tau_0},\frac{\tau_s}{\tau_0}) 
\end{equation}
where $\Psi^{eq}$ and $\Psi^{neq}$ encode the effects of turbulent and 
kinetic fluctuations, respectively.
 
It is now instructive to inspect whether the corresponding 
expressions derived from a RG treatment of the Navier stokes 
equations i) fit in the above expressions, and if so,
ii) whether the "universal" functions $\Psi^{eq}$ and $\Psi^{neq}$ are the same.  
   
The Yakhot-Orszag expression of the renormalized relaxation time derived
from a RG treatment of the Navier-Stokes equations, reads as follows \cite{YO}:
\begin{equation}
\label{YO}
r_{YO}-1=C_{YO} \frac{k}{T} \frac{\tau_l/\tau_0}{\sqrt{1+\eta^2}}
\end{equation}
where we have set $\eta=\tau_s/\tau_l = kS/\epsilon$, and the numerical 
constant is $C_{YO} \sim 0.08$. 

Under the assumption $\tau_{tur}/\tau_0 \gg 1$, the non-equilibrium component of the 
YO expression is exactly the same as the kinetic one, equation (\ref{TAUTUR}). 

As to the equilibrium component, from (\ref{YO}) one reads off:
$$
\delta F^{eq}/F^{eq} = \frac{C_{YO} k/T}{1-C_{YO}k/T} 
$$
In Fig 2 we compare the YO expression above with both 
kinetic expressions (\ref{RRT0}) (with $C_1=C_{YO}=0.08$) 
and (\ref{RRT}) for two different values of $\theta \equiv \tau_{tur}/\tau_0 = 10, 100$.

The figure shows that while all kinetic expressions provide a satisfactory
agreement with the YO formulation for $k/T$ below about $0.1$, above such value
the kinetic formulations predict a significantly larger relaxation time.

\begin{figure}
\label{FIG2}
\includegraphics[scale=0.3]{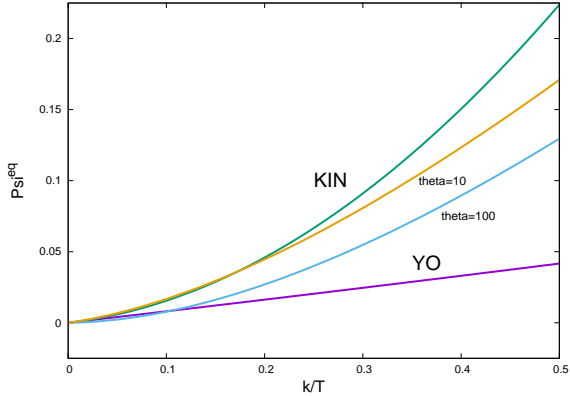}
\caption{The equilibrium contribution to the renormalized relaxation rate: 
Yakhot-Orszag (YO), present derivation (KIN), namely expression \ref{RRT0} and
\ref{RRT} with $\theta=10,100$}.
\end{figure}

\section{Conclusion}

We have derived an ab-initio kinetic expression for the renormalized 
relaxation rate as a function of the dimensionless ratios 
$k/T$ (turbulent Mach number) and turbulent
time scales $\tau_l$ and $1/S$.

The kinetic expression shows strong similarity with the Yakhot-Orsag
expression, with a much larger quadratic term in the parameter $k/T$.
In the range $k/T<0.1$, the numerical values are comparable, but for larger values
the kinetic relaxation time significantly exceeds the YO value.   
It would be interesting to explore the effect of the new 
expression (\ref{RRT}) in hydrokinetic simulations of turbulent flows.

\section{Acknowledgments}

The author owes a huge debt of knowledge to 
Hudong Chen, Viktor Yakhot and to the late 
SA Orszag, a wonderful master and a much missed friend. 

This paper was prepared on the occasion of the Simons Symposium
"Universality: Turbulence across vast scales".
The author is grateful to the Simons Foundation for 
financial support and great hospitality and to
David Spergel and Phil Mocz for very stimulating discussions.

This research has also received funding from the European Research Council, 
under the European Union's Horizon 2020 Framework Programme (No. FP/2014-
2020)/ERC Grant Agreement No. 739964 (COPMAT).


{}

\end{document}